\documentclass{article}
\usepackage{graphicx}
\usepackage{psfrag}

\psfrag{Ni/N}[tr][][0.8][270]{$\frac{n_{i}}{N}$}
\psfrag{sqrt(s),MeV}[t][][0.6][0]{$\sqrt{s} \mbox{,~MeV}$}
\psfrag{sigma,nb}[br][][0.8][0]{$\sigma\mbox{,~nb}$}
\psfrag{IL, nb-1}[br][][0.8][0]{$IL\mbox{,~nb}^{-1}$}
\psfrag{mgg,MeV}[t][][0.8][0]{$m_{\gamma\gamma}$\mbox{,~MeV}}
\psfrag{x23g}[t][][0.8][0]{$\chi^{2}_{3\gamma}$}
\psfrag{ND}[l][l][0.4][0]{\mbox{ND}}
\psfrag{SND99}[l][l][0.4][0]{\mbox{SND99}}
\psfrag{this study}[l][l][0.5][0]{\mbox{this work}}
\psfrag{sq(s)=782 MeV}[l][l][0.6][0]{$\sqrt{s}=782$\mbox{~MeV}}
\psfrag{Er, MeV}[l][l][0.6][0]{$E_{r}$,\mbox{~MeV}}
\psfrag{e(s(s),Er)/e(s(s),0)}[l][l][0.5][0]{$\varepsilon_{r}(\sqrt{s},E_{r})/\varepsilon_{r}(\sqrt{s},0)$}
\psfrag{dEr=64 MeV}[l][l][0.6][0]{$\delta{}E_{r}\sim{}64.4$\mbox{~MeV}}
\psfrag{dEr, MeV}[l][l][0.6][0]{$\delta{}E_{r}$\mbox{,~MeV}}
\psfrag{sq(s), MeV}[l][l][0.8][0]{$\sqrt{s}$\mbox{,~MeV}}
\psfrag{sqrt(s), MeV}[l][l][0.8][0]{$\sqrt{s}$\mbox{,~MeV}}
\psfrag{eff(s)}[l][l][0.8][0]{$\varepsilon{}(\sqrt{s})$}
\psfrag{beta(s)}[l][l][0.8][0]{$\beta{}(\sqrt{s})$}
\psfrag{tet1,dg}[l][l][0.8][0]{$\theta_{1}^{\circ}$}
\psfrag{tet3,dg}[l][l][0.8][0]{$\theta_{3}^{\circ}$}
\psfrag{Ep3,MeV}[l][l][0.8][0]{$E_{\gamma,3}$, MeV}
\psfrag{eLay3/eLay12}[l][l][0.8][0]{$\frac{E_{l3}}{E_{l1}+E_{l2}}$}
\psfrag{Etot/sqrt(s)}[l][l][0.8][0]{$\frac{E_{tot}}{\sqrt{s}}$}
\psfrag{P/sqrt(s)}[l][l][0.8][0]{$\frac{P_{tot}}{\sqrt{s}}$}

\psfrag{Grpg, kev}[r][r][0.5][0]{$\Gamma_{\rho\rightarrow\pi^{0}\gamma}$, keV}
\psfrag{Gopg, kev}[r][r][0.5][0]{$\Gamma_{\omega\rightarrow\pi^{0}\gamma}$, keV}
\psfrag{Gppg, kev}[r][r][0.5][0]{$\Gamma_{\phi\rightarrow\pi^{0}\gamma}$, keV}

\psfrag{min mij}[l][l][0.8][0]{$m_{3}$, MeV}
\psfrag{max mij}[l][l][0.8][0]{$m_{2}$, MeV}
\psfrag{ph peak}[c][c][0.8][0]{$\phi$ peak}
\psfrag{om peak}[c][c][0.8][0]{$\omega$ peak}
\psfrag{cross, nb}[l][l][0.8][0]{$\sigma$, nb}

\begin{document}

\title{Study of the \( e^{+}e^{-}\rightarrow \pi ^{0}\gamma  \) process
in the energy range \( 0.60-1.06 \)~GeV.}


\author{M.N.Achasov,
 K.I.Beloborodov,
 A.V.Berdugin,\\
 A.G.Bogdanchikov,
 A.V.Bozhenok,
 A.D.Bukin,
 D.A.Bukin,\\
 A.V.Vasiljev,
 T.V.Dimova,
 V.P.Druzhinin,
 V.B.Golubev,\\
 {\em A.A.Korol}\thanks{Talk given at the X International Conference on Hadron Spectroskopy, Aschaffenburg, Germany, August 31 - September 6, 2003 \protect \\ {\em e-mail: korol@inp.nsk.su}},
 S.V.Koshuba,
 A.P.Lysenko,
 E.V.Pakhtusova,\\
 S.I.Serednyakov,
 Yu.M.Shatunov,
 V.A.Sidorov,
 Z.K.Silagadze,\\
 A.N.Skrinsky,
 Yu.V.Usov\\*[0.5cm]
\em Budker Institute of Nuclear Physics \\ \em Siberian Branch of the Russian Academy of Sciences}

\date{September 2, 2003}

\maketitle

\begin{abstract}
The process \( e^{+}e^{-}\rightarrow \pi ^{0}\gamma  \) in the energy
range \( \sqrt{s}=0.60-1.06 \)~GeV was studied at VEPP-2M collider
with SND detector using \( \sim 14\, pb^{-1} \) of integrated luminosity.
Data were analyzed in the framework of the vector meson dominance
model. Preliminary results on obtained cross-section and parameters
of \( \rho ^{0},\omega ,\phi \rightarrow \pi ^{0}\gamma  \) radiative
decays are presented.
\end{abstract}

\section{Introduction}

In the vector meson dominance (VMD) model the \( e^{+}e^{-}\rightarrow \pi ^{0}\gamma  \)
process is considered as a transition \( e^{+}e^{-}\rightarrow \rho ^{0},\, \omega ,\, \phi \rightarrow \pi ^{0}\gamma  \)
. Further the decay parameters can be related with \( \omega ,\, \phi \rightarrow \rho ^{0}\pi ^{0} \)
and \( \pi ^{0}\rightarrow 2\gamma  \) \cite{cite:Gellmann:1962}.
Another approach to the decays is a non-relativistic quark model (NQM).
In this model vector mesons considered as composite from valent quark
with codirected spins, while pseudoscalar mesons with opposite directed
spins. The decay may be explained as spin overturn with photon emission
(magneto-dipole transition)\cite{cite:Geffen:1980}. The study of
such decays is important for understanding light mesons structure
and tests of strong interactions at low energies.

\section{Experiment}

The experiment was made with SND detector \cite{cite:Achasov:1999ju}
at VEPP-2M collider\cite{cite:Skrinsky:1995}. The SND detector
consists of tracking system, electromagnetic calorimeter and muon
veto system. The principal part of the detector is an electromagnetic
calorimeter. Full thickness of calorimeter for particles originating from the detector
center is \( 13.4\, X_{0} \), total solid angle is \( 90\%\cdot 4\pi  \).
Calorimeter energy resolution for photons is \( \frac{\sigma _{E}}{E}\approx \frac{4.2\%}{E(GeV)^{1/4}} \),
its angle resolution is \( \sigma _{\varphi ,\theta }\approx \frac{0.82^{\circ }}{\sqrt{E(GeV)}}\oplus 0.63^{\circ } \)
\cite{cite:Achasov:1998ep}.

Presented work is an extension of the studies published in Ref.\cite{cite:Achasov:phip0g,cite:Achasov:2003ed}
but using additional data. The data were collected in experiments
in 1998 and 2000 \cite{cite:Achasov:2001yz}:
\begin{itemize}
\item PHI98: energy range \( 0.98-1.06 \)~GeV, integrated luminosity \( 7.83\, \, pb^{-1} \)
(\( \sim 1.2\cdot 10^{7} \) \( \phi  \)-mesons)
\item OME00: energy range \( 0.60-0.97 \)~GeV, integrated luminosity \( 5.93\, \, pb^{-1} \)
(\( \sim 2.5\cdot 10^{6} \) \( \omega  \)-mesons)
\end{itemize}

\section{Analysis}

The process \( e^{+}e^{-}\rightarrow \pi ^{0}\gamma  \) was studied
in the \( 3\gamma  \) final state. Main background sources are \( e^{+}e^{-}\rightarrow 3\gamma  \)
(QED origin); \( e^{+}e^{-}\rightarrow 2\gamma  \) misidentified
because of additional clusters from machine background; \( e^{+}e^{-}\rightarrow \eta \gamma  \),
and cosmic background. For luminosity calculation events of the \( e^{+}e^{-}\rightarrow 2\gamma  \)
process were used.

The events were accepted by first level trigger which allows two or
more clusters in calorimeter, no signal in either tracking or muon
system, energy deposition in calorimeter greater then some threshold
(which varied but never was greater than \( 0.4\sqrt{s} \)).

Preliminary selection included following cuts: no charged tracks,
number of clusters in calorimeter \( N_{np}\geq 3 \), calorimeter
energy deposition \( E_{cal}\geq 0.65\sqrt{s} \), calorimeter
total momentum \( P_{cal}\leq 0.3\sqrt{s} \), polar angles of
two most energetic clusters \( 36^{\circ }\leq \theta _{1,2}\leq 144^{\circ } \),
polar angle of the next by energy cluster \( 27^{\circ }\leq \theta _{3}\leq 153^{\circ } \),
its energy deposition \( E_{cal\, 3}\geq 0.1\sqrt{s} \).

Kinematic fit with energy and momentum conservation constraints was
applied to selected events. It improved \( \pi ^{0} \) mass resolution
from \( \sigma _{m_{\gamma \gamma }}=11.2 \)~MeV to \( \sigma _{m_{\gamma \gamma }}=8.6 \)~MeV.
Cut on the fit quality parameter \( \chi _{3\gamma }^{2}<20 \) was
then applied. Selected events were subdivided into two classes: \( 108 \)~MeV\( \leq m_{\gamma \gamma }\leq 162 \)~MeV
(class A) and the rest (class B). Class B events were used to cross-check
our understanding of background contribution and estimate related systematic
uncertainty. For class A \( \sim 7\cdot 10^{4} \) events were selected.

For luminosity calculation events were selected (class C) with no
charged tracks, two or more clusters with calorimeter energy deposition
\( E_{cal\, 1,2}\geq 0.3\sqrt{s} \), polar angles \( 36^{\circ }\leq \theta _{1,2}\leq 144^{\circ } \),
acollinearity \( \Delta \varphi _{12}\leq 10^{\circ } \), \( \Delta \theta _{12}\leq 25^{\circ } \).
Events satisfying preliminary selection cuts for A and B classes were
excluded. It is necessary to note significant contribution of \( e^{+}e^{-}\rightarrow \pi ^{0}\gamma  \)
events (up to \( 10\% \) at \( \omega  \) peak) to this class.

Cross section of the studied process was parametrized with usual
variable width Breit-Wigner description \cite{cite:Achasov:freshlook}.
More details published in Ref.\cite{cite:Achasov:2003ed}.
Cross sections \(\sigma _{V\pi ^{0}\gamma }\) of the \( e^{+}e^{-}\rightarrow V\rightarrow \pi ^{0}\gamma  \)
transitions at resonances mass (\(V=\rho ^{0},\omega ,\phi\)) were
approximation parameters. Other parameters considered were relative
phases \(\varphi_{\rho\omega}\) and  \(\varphi_{\phi\omega}\), and
constant real amplitude \(a_{\pi^{0}\gamma}\) taking into account possible
contribution from higher resonances decays. Masses and
widths of \( \omega, \, \rho ^{0}, \, \phi  \)-mesons and branching
ratios of \( \omega  \) and \( \phi  \) major decay modes were taken
from previous SND measurements \cite{cite:Achasov:pi3},
other external parameters were taken from review of particle physics
\cite{cite:Hagiwara:2002fs}.

Visible cross section of \( e^{+}e\rightarrow \pi ^{0}\gamma  \)
was calculated taking into account
radiative correction \cite{cite:Kuraev:radcor} and dependency on registration
efficiency from the energy of initial state radiation photons.
Registration efficiency for all processes was calculated with Monte Carlo
simulation.

Cross section of QED \( e^{+}e^{-}\rightarrow 2\gamma  \) was calculated
using Ref.\cite{cite:Baier:1981kx}. Cross section for \( e^{+}e^{-}\rightarrow 3\gamma  \)
process was calculated from tree-level diagram corrected using Ref.\cite{cite:Kuraev:th3grc}.
Cosmic background events fraction was estimated from the data collected
without beams.

Cross section approximation was done using maximum likelihood method.
Classes A, B, C were approximated at the same fit.
Integrated luminosity was recalculated at the every minimization step.

Four models were used for approximation: (1) no \( \rho  \) meson
decay contribution; (2) including \( \rho  \) meson contribution,
with fixed \( \rho -\omega  \) relative phase; (3) including \( \rho  \)
meson contribution, \(\varphi_{\rho\omega}\) estimated from 
electromagnetic \( \rho -\omega  \) mixing 
\cite{cite:Achasov:1992ku,cite:OConnell:mixing};
(4) including \( \rho  \) meson contribution, with electromagnetic
\( \rho -\omega  \) mixing fixed using \( B_{\omega \rightarrow 2\pi } \),
and constant real amplitude.

\section{Results and discussion}

The approximation results are summarized in the Table~\ref{tab:result}. 
\begin{table}

\caption{\label{tab:result}Approximation results}
\begin{tabular}{|ccccc|}
\hline 
No.&
1&
2&
3&
4\\
\hline
\( \chi ^{2}/N \)&
\( 235/56 \)&
\( 51/54 \)&
\( 51/54 \)&
\( 51/53 \)\\
\( \sigma _{\omega \pi ^{0}\gamma } \), nb&
\( 174.1\pm 0.9 \)&
\( 151.9\pm 1.5 \)&
\( 152.1\pm 1.6 \)&
\( 152.4\pm 1.6 \)\\
\( \sigma _{\rho ^{0}\pi ^{0}\gamma } \), nb&
\( 0 \)&
\( 0.59\pm 0.07 \)&
\( 0.58\pm 0.07 \)&
\( 0.56\pm 0.07 \)\\
\( \sigma _{\phi \pi ^{0}\gamma } \), nb&
\( 6.11\pm 0.17 \)&
\( 5.58\pm 0.29 \)&
\( 5.73\pm 0.31 \)&
\( 5.64\pm 0.37 \)\\
\( \scriptstyle \varphi _{\rho \omega },^{\circ } \)&
&
\( -9.3\pm 3.3 \)&
\( -9.3\pm 3.3 \)%
\footnotemark&
\( -12.6\pm 1.1 \)%
\footnotemark\\
\( \scriptstyle \varphi _{\phi \omega },^{\circ } \)&
\( 169\pm 8 \)&
\( 151\pm 9 \)&
\( 160\pm 9 \)&
\( 160\pm 10 \)\\
\( {\scriptstyle \mathbf{Re}}\, a_{\pi ^{0}\gamma } \),nb\( ^{\frac{1}{2}} \)&
\( 0 \)&
\( 0 \)&
\( 0 \)&
\( -0.05\pm 0.07 \)\\
\hline
\end{tabular}

\end{table}
\addtocounter{footnote}{-1}
\footnotetext{estimated from \(\rho -\omega\) mixing}
\stepcounter{footnote}
\footnotetext{estimated from \( \rho -\omega  \) mixing, \(\Pi _{\rho \omega }=-3676\pm 303 \)~MeV\( ^{2} \) fixed using \( B_{\omega \rightarrow 2\pi} \)}

\begin{figure}
\resizebox*{0.8\textwidth}{!}{\includegraphics{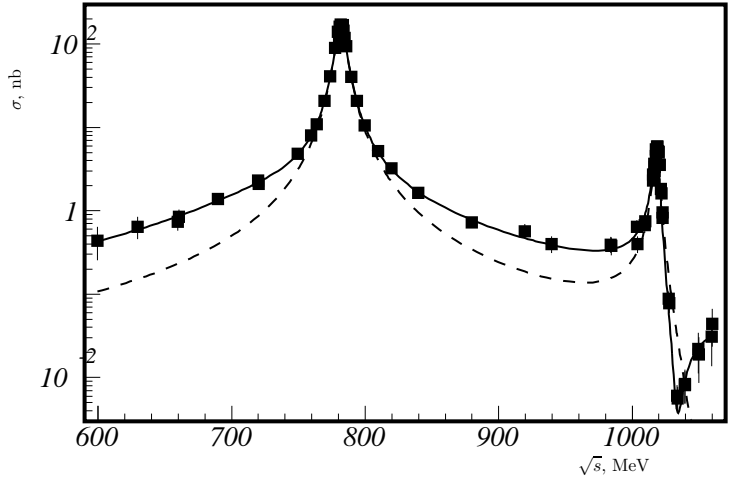}}

\caption{\label{fig:cross}Cross section of the \protect\( e^{+}e^{-}\rightarrow \pi ^{0}\gamma \protect \)
process. Dots represent experimental data, solid curve depicts approximated
theoretical form with \protect\( \rho ^{0}\rightarrow \pi ^{0}\gamma \protect \)
(model 2), dashed curve --- without \protect\( \rho ^{0}\protect \)
contribution (model 1).}
\end{figure}

Model (1) contradicts to experimental data, models (2)--(4) describe
data equally well. Obtained cross section is shown on Fig.~\ref{fig:cross}. 
 Main sources of systematic errors of cross section are uncertainty
in luminosity determination (\( 2\% \)), in registration efficiency
(\( 2.5\% \)), QED and \( e^{+}e^{-}\rightarrow \eta \gamma  \)
background subtraction (\( 0.5-20\% \) depends on energy point).
Cosmic background contribution was found negligible (\( <0.1\% \)).
Model and external parameters uncertainty also give contribution to
systematic errors of cross section parameters. 

Cross section parameters were found as:\begin{eqnarray*}
\sigma _{\rho ^{0}\pi ^{0}\gamma } & = & (0.59\pm 0.07\pm 0.06)\, nb\\
\sigma _{\omega \pi ^{0}\gamma } & = & (151.9\pm 1.5\pm 4.6)\, nb\\
\sigma _{\phi \pi ^{0}\gamma } & = & (5.58\pm 0.29\pm 0.29)\, nb\\
\varphi _{\rho \omega } & = & -9.3^{\circ }\pm 3.3^{\circ }\pm 2.4^{\circ }\\
\varphi _{\phi \omega } & = & 151^{\circ }\pm 9^{\circ }\pm 11^{\circ }
\end{eqnarray*}
and branching ratios of \( \rho ^{0},\, \omega ,\, \phi \rightarrow \pi ^{0}\gamma  \)
decays:

\begin{eqnarray*}
B_{\omega \rightarrow \pi ^{0}\gamma } & = & (8.45\pm 0.09\pm 0.25)\, \%\\
B_{\rho ^{0}\rightarrow \pi ^{0}\gamma } & = & (5.32\pm 0.63\pm 0.50)\cdot 10^{-4}\\
B_{\phi \rightarrow \pi ^{0}\gamma } & = & (1.34\pm 0.07\pm 0.07)\cdot 10^{-3}
\end{eqnarray*}
The measured decays partial widths and previous experimental data
are summarized in the Table~\ref{tab:comparison}.
\begin{table}

\caption{\label{tab:comparison}Comparison with previous experimental data.}

\begin{tabular}{|llll|}
\hline 
\textbf{Source}&
 \( \Gamma _{\omega } \), keV&
 \( \Gamma _{\rho } \), keV&
\( \Gamma _{\phi }, \) keV\\
\hline 
This work&
\( 733\pm 7\pm 22 \)&
\( 80.4\pm 9.7\pm 7.6 \)&
\( 5.98\pm 0.31\pm 0.31 \) \\
SND (2000,2003) \cite{cite:Achasov:phip0g,cite:Achasov:2003ed}&
\( 788\pm 12\pm 27 \)&
\( 77\pm 17\pm 11 \)&
\( 5.40\pm 0.16_{-0.40}^{+0.43} \)\\
PDG2002 \cite{cite:Hagiwara:2002fs}&
\( 747\pm 41 \)&
\( 121\pm 31 \)&
\( 5.28\pm 0.43 \)\\
\( \omega \rightarrow  \)\emph{neutrals} (PDG2002 \cite{cite:Hagiwara:2002fs})&
\( 767\pm 79 \) &
&
\\
\( \rho ^{\pm }\rightarrow \pi ^{\pm }\gamma  \)
(PDG2002\cite{cite:Hagiwara:2002fs})&
&
\( 68\pm 7 \)&
\\
\hline
\end{tabular}
\end{table}
Measured parameters agree with previous experimental data and mostly
compatible with phenomenological expectations (see e.g. tables in
Refs.\cite{cite:Achasov:phip0g,cite:Achasov:2003ed}). Relative phase of
\( \rho  \) and \( \omega  \) decays may be attributed to \( \rho -\omega  \)
mixing (\( -12.6^{\circ }\pm 1.1^{\circ } \) from \( B_{\omega \rightarrow 2\pi } \)).
Partial width ratios \( \frac{\Gamma _{\omega \rightarrow \pi ^{0}\gamma }}{\Gamma _{\rho ^{0}\rightarrow \pi ^{0}\gamma }}=9.11\pm 1.10\pm 0.90 \)
and \( \frac{\Gamma _{\phi \rightarrow \pi ^{0}\gamma }}{\Gamma _{\omega \rightarrow \pi ^{0}\gamma }}=(8.16\pm 0.43\pm 0.48)\cdot 10^{-3} \)
are compatible with SU(3) predictions.

\section{Conclusion}

The cross section of the \( e^{+}e^{-}\rightarrow \pi ^{0}\gamma  \)
process is measured in the c.m. energy region \( \sqrt{s}=0.60-1.06 \)~GeV
using the data obtained by the SND detector at the VEPP-2M collider.
Total integrated luminosity used is \( \sim 14\, pb^{-1} \), \( 7\cdot 10^{4} \)
events were selected for analysis. 

Data were analyzed in the framework of VMD model, parameters of decays
\( \rho ^{0},\omega ,\phi \rightarrow \pi ^{0}\gamma  \) were obtained.
Parameters of decays agree with previous measurements and phenomenological
estimations. Partial width of the \( \rho ^{0}\rightarrow \pi ^{0}\gamma  \)
decay is close to the \( \rho ^{\pm }\rightarrow \pi ^{\pm }\gamma  \)
one. The relative phase for \( \rho ^{0} \), \( \omega  \) decays
can be attributed to electromagnetic \( \rho -\omega  \) mixing. 

\section{Acknowledgments}

One of the authors thanks organizers of the conference for the invitation and
support. This work was supported in part by Presidential Grant
1335.2003.2 for support of Leading Scientific Schools and by Russian Science
Support Foundation.


\end{document}